\documentclass[twocolumn,prb, superscriptaddress]{revtex4-2}
\usepackage[usenames,dvipsnames]{color}

\usepackage{comment}
\usepackage{graphicx}
\usepackage[utf8x,utf8]{inputenc}
\usepackage{lineno}
\usepackage{color}
\usepackage{hyperref}
\usepackage{array}
\usepackage{amsmath}
\usepackage{braket}
\usepackage{caption}
\usepackage{subcaption}
\usepackage{setspace}
\usepackage[section]{placeins} 
\usepackage{multirow}
\usepackage{booktabs}
\usepackage{xcolor}
\usepackage{amssymb}
\usepackage{cancel}
\usepackage{float}
\usepackage{soul}

\usepackage[]{todonotes}

\begin{document}

\title{Lifshitz transition and triplet $p$-wave pairing from the induced ferromagnetic plaquette via spin differentiated nonlocal interaction}

\author{Rayan Farid}
\affiliation{Department of Physics and Physical Oceanography, Memorial University of Newfoundland and Labrador, St. John's, Newfoundland \& Labrador, Canada A1B 3X7} 

\author{Daria Gazizova }
\affiliation{Department of Physics and Physical Oceanography, Memorial University of Newfoundland and Labrador, St. John's, Newfoundland \& Labrador, Canada A1B 3X7} 
\author{B. D. E. McNiven}
\affiliation{Department of Electrical and Computer Engineering, Memorial University of Newfoundland and Labrador, St. John's, Newfoundland \& Labrador, Canada A1C 5S7} 

\author{J. P. F. LeBlanc}
\email{jleblanc@mun.ca}
\affiliation{Department of Physics and Physical Oceanography, Memorial University of Newfoundland and Labrador, St. John's, Newfoundland \& Labrador, Canada A1B 3X7} 

\date{\today}
\begin{abstract}
We study the two-dimensional extended Hubbard model on a square lattice and incorporate spin-differentiated nearest neighbor (NN) interactions where the equal-spin ($V_{uu}$) and unequal-spin ($V_{ud}$) terms are independently tuned parameters. We compute single-particle excitations as well as static spin and pairing susceptibilities perturbatively up to the fourth order within the thermodynamic limit and at a finite fixed temperature. By explicitly encoding a ferromagnetic-like NN interaction ($V_{uu} < V_{ud}$), we induce a competition among the uniform $q = (0,0)$, collinear $q = (\pi,0)$, and staggered $q = (\pi,\pi)$ spin excitations.  This results in the formation of short-ranged $2\times 2$ ferromagnetic plaquettes arranged in staggered  or striped patterns. Kinetic frustration in hopping, both within and between these plaquettes, manifests in single-particle properties, resulting in a reduction of bandwidth and ultimately triggering a Lifshitz transition to quasi-one-dimensional bands. Furthermore, an attractive effective interaction within the localized ferromagnetic plaquette results in the emergence of equal-spin triplet $p$-wave pairing. We demonstrate that sufficiently strong magnetic fluctuations, even at finite length scales, can significantly influence single-particle and pairing properties without breaking translational symmetry. Our approach provides a novel pathway to realize a variety of rich magnetic phases and Fermi surface reconstruction driven by interactions in the absence of explicit geometric frustration.
\end{abstract}
\maketitle

\section{Introduction}
\label{sec:intro}
Charge and spin density waves are ubiquitous features of strongly correlated materials, prominently observed in high-$T_c$ cuprates and iron chalcogenides \cite{comin2016,Tranquada2020,Cruz2008,Stewart2011}. In particular, growing evidence suggests that a delicate interplay between spin and charge stripe order is involved in either facilitating or competing with $d$-wave pairing in doped cuprates \cite{Neto2013,Neto2018,Guguchia2023,Keimer2015,Jiang2022}. In superconducting iron chalcogenides, the parent phase exhibits more exotic spin ordering, including colinear, pair-checkerboard, and plaquette antiferromagnetic orders \cite{Li2010,Machida2012,Cao2015,Taylor2013,Tam2019}, which have been shown to arise from excitations in different spin modes and multiorbital effects \cite{Glasbrenner2015,Ducatman2012,Chubukov,Fernandes_2017}. The prevailing consensus is that understanding the microscopic mechanism behind charge and magnetic fluctuations will shed further light on superconductivity \cite{Scalapino2012}. This pursuit has led to the development of various models
and numerical techniques, with the 2D Hubbard model
taking a paradigmatic role  due to its ability to capture a wide range of correlated phenomena despite its apparent simplicity\cite{Hubbard,Arovas2022,mingpu:2022,leblanc:benchmark}. In fact, it is established that the ground state of
Hubbard model antiferromagnetic at half-filling \cite{Hirsch1989}. The 
 doped regime exhibits both commensurate and
incommensurate spin orders and charge stripes, thought to be relevant for cuprates \cite{Zheng2017,Huang2017,Mai2023}.

In contrast, hosting ferromagnetic orders is difficult in the single band Hubbard model and occurs only under specific conditions. For example, Nagaoka rigorously proved that in the limit of infinite Hubbard interaction ($U\to\infty$)  the presence of a single hole renders the ground-state of the Hubbard model to be ferromagnetic \cite{Nagaoka1966,Tasaki1989}. However, Nagaoka's stringent condition limits its applicability in realistic systems \cite{Ilya2017,Hanisch,Kollar1996}. Consequently, alternative routes, such as flat-band systems or modified Hubbard model with anisotropic hopping have been considered to induce ferromagnetism \cite{Mielke1991,Samajdar1,Samajdar2}. In particular, kagome materials have attracted a lot of interest recently due to their ability to hold robust flat-band ferromagnetism driven by frustration \cite{Lin2018,Samanta2024}. Furthermore, to model the parent phase of iron chalcogenides, the minimal single band model involves mapping the half-filled Hubbard model to a Heisenberg spin model and then tuning the nearest- and next-nearest exchange coupling strengths to create relevant phases \cite{Hu2012,Glasbrenner2015}. A recent study has also tuned the nearest- and next-nearest hopping parameters to frustrate the lattice geometry at finite $U$ values which then induces magnetic phases in the ground state\cite{Ruan2021,Fernandes_2017}. What would be ideal is a unified yet low-energy single-band model driven by interaction, capable of capturing and hosting a variety of rich magnetic phases at finite temperatures, without invoking explicit frustration or additional orbital degrees of freedom.

In this work, we address this issue by incorporating a nearest-neighbor (NN) interaction,  $V_{\sigma\sigma^\prime}$, into the single-band  Hubbard model on an infinite square lattice. Our approach deviates from the standard extended Hubbard model by allowing the magnitude and signs of the equal-spin, $V_{uu}$, and unequal-spin, $V_{ud}$, NN interaction terms to be independently assigned. Prior studies with a non-local repulsive interaction are based on the assumption that the interaction is not spin dependent, and that this is the result of an effective non-local density-density interaction \cite{Paki201,Peng2023,Hanna2021}. However, it was shown that the fully renormalized local $U$ interaction eventually gives rise to a non-local but spin-dependent $V$ interaction even in the weak coupling limit \cite{gazon2023}. This motivates us to study the $V_{uu} < V_{ud}$ case that favors the alignment between equal spins in neighboring sites, thereby promoting competition between ferromagnetic and antiferromagnetic fluctuations without inducing geometric frustration in the lattice. 

For conceptual simplicity, we study the 2D Hubbard model at half-filling with only NN hopping $t$.  Aided by Algorithmic Matsubara Integration (AMI) \cite{AMI2019,AMI2}, we compute single- and two-particle response functions from direct diagrammatic perturbation theory, up to the fourth order within the thermodynamic limit at a fixed temperature of $\beta t=5$. Due to recent advances in AMI, it is possible to disentangle the powers of the $U$, $V_{uu}$, and $V_{ud}$ interactions symbolically from Feynman diagrams to obtain the multi-indexed coefficients, yielding full access to the interaction space once the bare coefficients are computed \cite{Me2024}. With this computational advantage, we compute the static spin susceptibility and identify a competition between two sets of magnetic fluctuations: a plaquette antiferromagnetic order and a doubly degenerate plaquette collinear stripe order along the lattice axes where the plaquette structure consists of the localized ferromagnetic domain. The kinetic frustration resulting from this competition manifests in single-particle properties, with a gradual reduction in bandwidth 
leading to the emergence of quasi-1D behavior near the bottom of the band and we will show that this triggers a Lifshitz transition.  We attribute this transition to the formation of additional poles in the zero-frequency quasiparticle dispersion.  In an infinite 2D system, translational symmetry cannot be spontaneously broken except at zero temperature due to the Mermin-Wagner theorem \cite{Mermin}. Nonetheless, our results demonstrate that finite range magnetic orders at a nominal non-zero temperature can affect single- and two-particle properties.  

Furthermore, within the ferromagnetic plaquette, the effective attractive interaction between NN aligned spins creates a potential for Cooper pair formation with $S_{z}=\pm1$. To demonstrate this, we compute the uniform pairing susceptibility for different pairing symmetry and find that triplet $p$-wave pairing with $S_{z}= \pm 1$ is the leading symmetry in our calculations. The equal-spin triplet pairing has so far remained elusive and has been explored  for its potential applications in superconducting spintronics due to its particular resistance to the pair breaking Zeeman effect or impurity scattering \cite{Eschrig_2015}. It is a potential candidate in certain uranium-based  superconductors \cite{Aoki_2019,Aoki_2022,Rosuel} and can be engineered in superconductor-ferromagnet heterostructures  via interfacial spin mixing or spin-rotation process \cite{Khaire,Diesch,Pal}. Thus, our model also serves as a minimal model under which equal spin pairing occurs  without the need for spin-orbit coupling, net global magnetization, or any interfacial effects.

\section{Model and Methods}
\subsection{Hubbard Hamiltonian}
The non-interacting Hubbard Hamiltonian of an infinite 2D square lattice with translational invariance is given by:
\begin{align}
    H_{K} =  -t \sum_{\langle ij \rangle, \sigma} c^{\dagger}_{i\sigma} c_{j\sigma} 
    - \mu \sum_{i\sigma} n_{i\sigma}.
\end{align}
Here, $c^{\dagger}_{i\sigma}$ ($c_{i\sigma}$) is the fermionic creation (annihilation) operator, $t$ is the hopping amplitude between NN sites $\langle ij \rangle$, $\{\sigma,\sigma^{\prime} \} \in \{\uparrow,\downarrow\}$ denote spin indices,  $\mu$ is the chemical potential which controls the filling, and $n_{i\sigma} = c^\dagger_{i\sigma} c_{i\sigma}$ is the density operator. The single-particle non-interacting dispersion of $H_{K}$ is 
\begin{equation}
\epsilon = -2t\big[ \cos(k_{x}) + \cos(k_{y})\big] - \mu. 
\end{equation}

 The interacting Hamiltonian $H_{U}$ consists of a local onsite Hubbard interaction $U$ between opposite spins and a nearest-neighbor interaction between equal ($V_{uu}$) and opposite ($V_{ud}$) spins, resulting in:
\begin{multline}
   H_{U}=\frac{U}{2} \sum_{i, \sigma \neq \sigma'} n_{i\sigma} n_{i\sigma'} 
    + \frac{V_{uu}}{2} \sum_{\langle ij \rangle, \sigma} n_{i\sigma} n_{j\sigma}\\ 
    + \frac{V_{ud}}{2} \sum_{\langle ij \rangle, \sigma \neq \sigma'} n_{i\sigma} n_{j\sigma'}. 
\end{multline}
With this, the total Hamiltonian is  the sum of the two components $H= H_{K}+H_{U}$. In the $V_{uu} > V_{ud}$ case, the model favors opposite-spin alignment between neighboring sites, which, in combination with $U$, further enhances fluctuations in a checkerboard antiferromagnetic (AFM) order. Conversely, the $V_{uu} < V_{ud}$ case promotes the alignment of equal spins between neighboring sites, potentially forming ferromagnetic domains that are expected to compete with the natural AFM tendencies in the half-filled model. We primarily focus on the latter in this study. Despite the spin-polarized nature of the interaction, the spin-degeneracy of the band is not lifted. This is due to the symmetry of equal and unequal spin interaction, such that the relation $G_{\uparrow\uparrow} = G_{\downarrow \downarrow}$ holds for the single particle propagator. We further parameterize interactions in $H_{U}$ into effective equal ($W_{uu}$) and unequal ($W_{ud}$) spin interactions. In the momentum space representation, they are defined as:
\begin{align}
    W_{uu}(q) &=  V_{uu}[2\cos(q_{x}) + 2\cos(q_{y})],\\
    W_{ud}(q) &= U\Big[1 + \frac{V_{ud}}{U}[2\cos(q_{x}) + 2\cos(q_{y})] \Big].
\end{align}
Using the $W_{uu}$ and  $W_{ud}$ interactions, we symbolically construct the diagrammatic expansions for the single-particle irreducible self-energy and the two-particle reducible particle-hole and particle-particle diagrams up to the fourth order \cite{Me2024}. At truncated fourth order, this generates a large set of diagrams: 571 for the self-energy, 4537 in the particle-hole channel, and 1982 in the particle-particle channel for both singlet and triplet pairing.

The parametrization of $ W_{ud}$  necessitates fixing the ratio $V_{ud}/U$. Once this is done, the amplitudes of $ V_{uu}$ and $ U$ interactions can be factored out from each diagram. The compiled diagrammatic sets are subsequently grouped based on the number of $ V_{uu}$ and $U$ interactions present in each diagram. From this, we obtain multi-indexed coefficients $a_{[i,j]}$ in powers of $U^{i}$ and $V_{uu}^{j}$, resulting in the truncated power series expansion:
\begin{align}
    \label{eq:coeff}
     O(U,V_{uu}) &= \sum a_{[i,j]}U^{i}V_{uu}^{i} \nonumber \\
     &=a_{[1,0]}U +  a_{[0,1]}V_{uu} + a_{[1,1]}UV_{uu} + a_{[2,0]}U^2 \nonumber \\
     & \quad + a_{[0,2]}V_{uu}^2 + a_{[3,0]}U^3 + a_{[0,3]}V_{uu}^3   \ldots 
\end{align}
such that $i+j\leq 4$ and the observables, $O(U,V_{uu})$, correspond to the single particle self-energy, the spin susceptibility or the pairing susceptibility studied in this work. Throughout, we set $t=1$ and operate in units of $t$. Moreover, we limit our calculations to a fixed ratio of $V_{ud}/U=0.2$, but note that $U$ and $V_{uu}$ can be changed without additional computational cost.

\subsection{Algorithmic  Matsubara Integration (AMI)}
AMI is a tool that provides the semi-analytical solution to any given Feynman integrand by symbolically resolving the Matsubara frequency via repeated application of the residue theorem \cite{AMI2019,elzab,burke2}. Once the temporal integrals are resolved, the momentum integrations are performed stochastically as continuous variables without discretization. This ensures that our results are always in the thermodynamic limit without the need for finite size extrapolations. Moreover, this approach offers the advantage of true analytical continuation of the external Matsubara frequency to the real axis via $i\omega \rightarrow \omega + i\gamma$ without requiring any ill posed numerical inversion technique \cite{AMI2,burke2}.   The only caveat is that the sharpness of spectral peaks depends on the choice of the finite regulator $\gamma$. In practice, a large and finite regulator is taken as  $\gamma = 0.1$$\sim$$0.2$ at the high temperature limit to ensure that calculations are not prohibitive numerically, without introducing large enough systematic bias to shift the qualitative features. However, when evaluating on the Matsubara axis no further considerations are required - such is the case for the zero bosonic matsubara frequency modes of the two particle susceptibility functions. These are evaluated directly from AMI.
\label{sec:model}
\section{Result and Discussion}
\subsection{Static spin susceptibility}
With the aim of identifying any emergent spin configurations within the model at half filling, we first identify the wavevector $\mathbf{q}$ dependence of magnetic excitations. This provides us with an initial insight into the possible magnetic fluctuations and will allow us to obtain the real-space fluctuations as well. To do so, the static, $i\Omega =0$, spin susceptibility is computed  diagrammatically in reciprocal space via the expression 
\begin{equation}
\chi_{s}(\mathbf{q})= \int^{\beta}_{0} d\tau \langle S_{z} (\mathbf{q},\tau) S_{z}(\mathbf{-q},0) \rangle e^{i\Omega \tau},
\end{equation}
where $S_{z}$ denotes the spin operator $\frac{1}{2}(n_{\uparrow}-n_{\downarrow})$. Both the reducible and irreducible particle hole diagrams are considered, without employing a resummation scheme to avoid any spurious divergence. We first consider the simplest  FM-like interaction by considering spin polarized case with equal magnitude but opposite sign ($V_{uu}/U=-0.2$ and $V_{ud}/U=0.2$). This serves as a controlled starting point where we expect to induce sufficiently strong FM fluctuations to compete with  AFM fluctuation associated with $(\pi,\pi)$ nesting at half filling.  Later, we present a more general $V_{uu}< V_{ud}$ case where the magnitude of $V_{uu}/U$ is varied with fixed repulsive $V_{ud}/U=0.2$. 

\begin{figure}[h]
    \centering
    \includegraphics[width=1\linewidth]{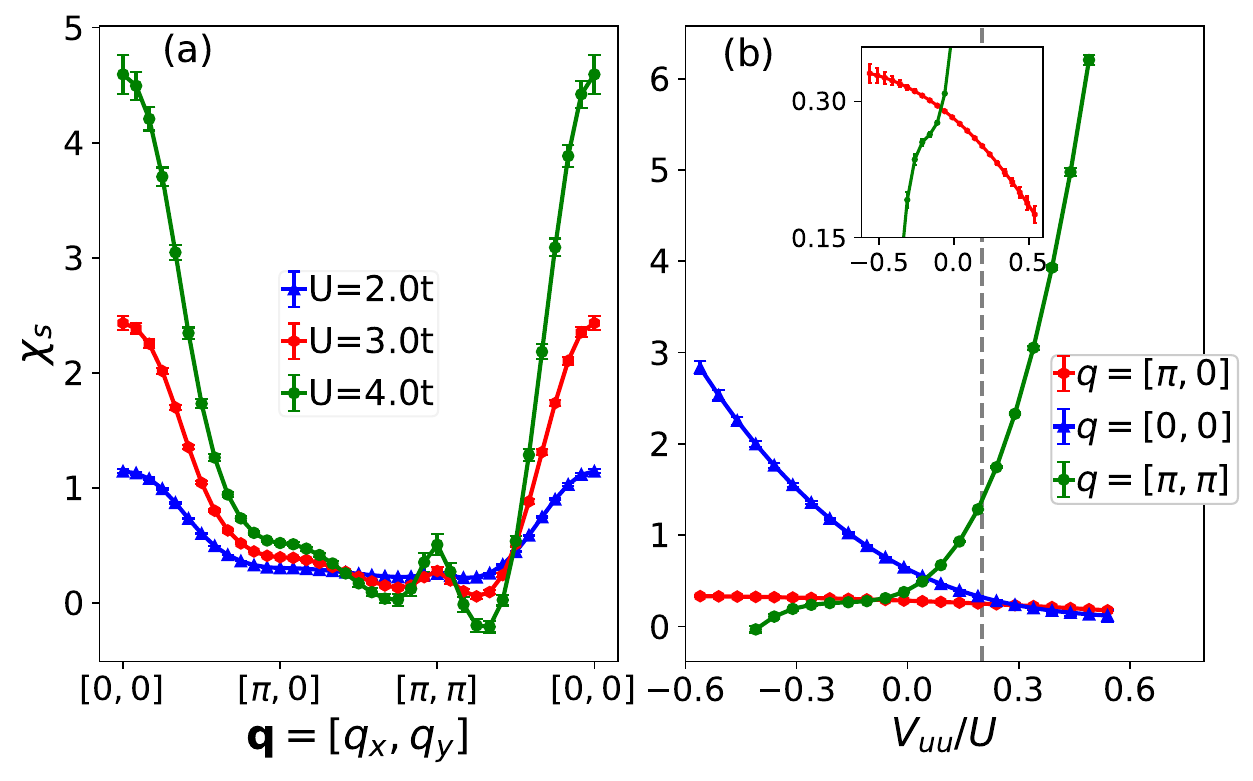}
    \caption{a) Static $\chi_{s}$(q) along a high symmetry momenta for $U/t=2,3$ and $4$ at $\beta=5$ and half filling. We fix the ratio $V_{uu}/U=-0.2$ and $V_{ud}/U=0.2$. b)  Three modes of $\chi_{s}$  plotted as function  of $V_{uu}/U$ with fixed $U/t=2$ and $V_{ud}/U=0.2$. The dashed grey vertical line shows where $V_{uu} = V_{ud}$.}
    \label{fig:spin_kcut}
\end{figure}

Fig.~\ref{fig:spin_kcut}(a) presents static  $\chi_{s}(q)$ along the high symmetry momentum cuts for interaction strengths of $U/t=2,3$ and $4$ with fixed ratio of $V_{uu}/U=-0.2$ and $V_{ud}/U=0.2$. Evidence of ferromagnetic (FM) and AFM fluctuations can be observed via the presence of two separate modes located at $q=[0,0]$ and $q=[\pi,\pi]$, respectively. The $q=[0,0]$ mode, representing uniform alignment of spin, has a significantly larger amplitude compared to the $q=[\pi,\pi]$ mode which corresponds to the staggered spin configuration. The amplitude of the two modes is positively correlated with the interaction strength $U$. Notably, the $q=[0,0]$ mode decays slowly along the direction $q=[q_{x},0]$ and shows a broad tail with an amplitude comparable to the $q=[\pi,\pi]$ mode. This has two implications. First, the slowly decaying $q=[0,0]$ mode is indicative of short ranged FM domain in real space. Second, the sizable tail extending to  $q=[\pi,0]$ indicates that the FM domain eventually decays to a fluctuating stripe structure along the lattice axes that may compete or coexist with checkerboard AFM structure.
\begin{figure}
    \centering
    \includegraphics[width=1\linewidth]{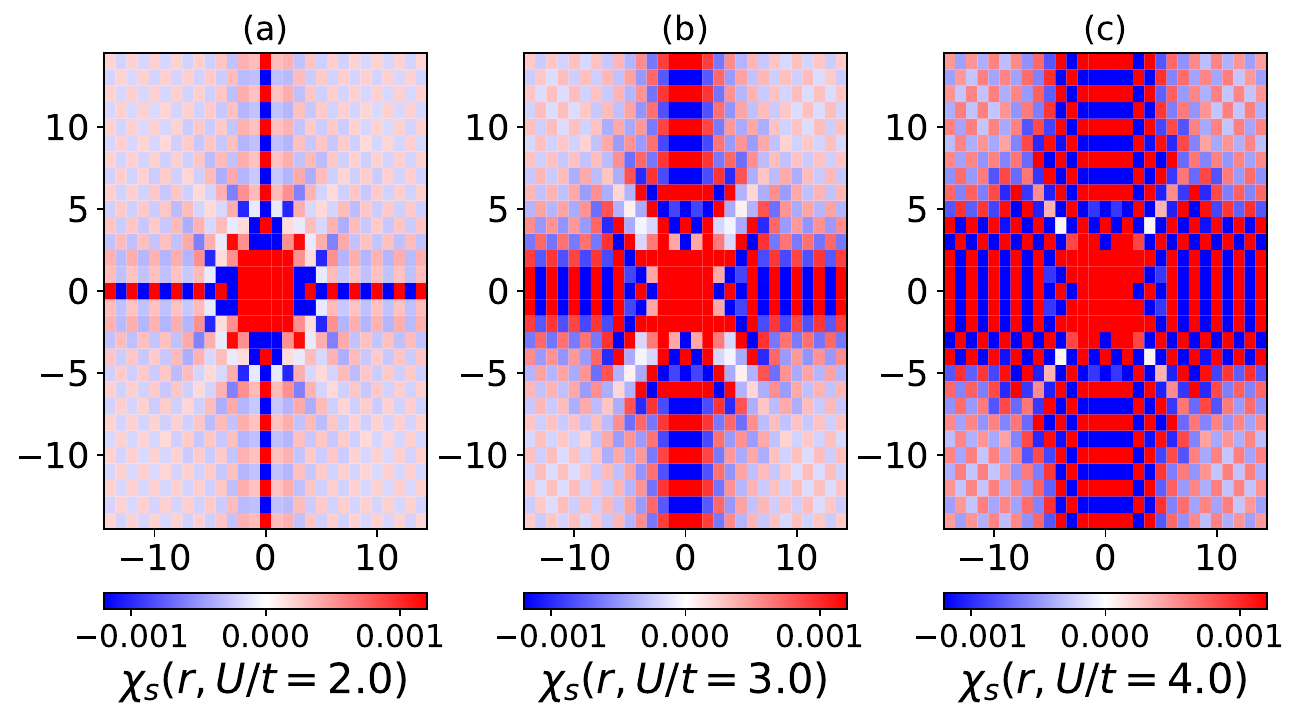}
    \includegraphics[width=1\linewidth]{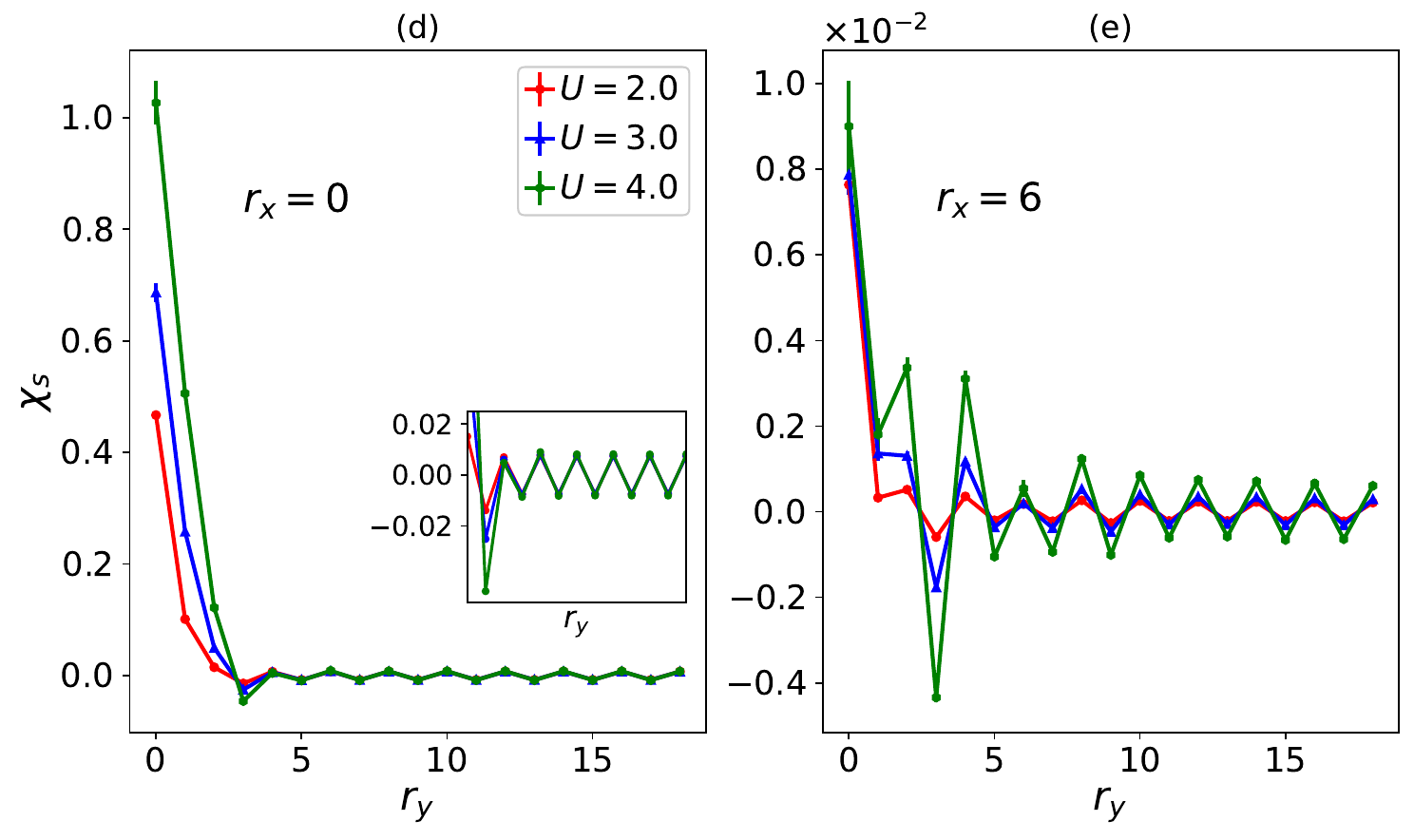}
     \caption{(a-c) False color plot of $\chi_s$ in real space with same parameter choice  as Fig.~\ref{fig:spin_kcut}(a). (d-e) vertical cuts of the color plot at $r_x=0$ and $r_x=6$, respectively. }
    \label{fig:FM_chi_r}
\end{figure}

In Fig.~\ref{fig:spin_kcut}(b), we focus on the key momenta at $q=[0,0]$, $q=[\pi,0]$ and $q=[\pi,\pi]$ and plot the amplitude of just these modes as a function of $V_{uu}/U$ where the value of $U/t=2$  with a fixed ratio of $V_{ud}/U=0.2$. Here, we assess the impact of varying $V_{uu}$ on the amplitude of these modes by considering two regimes, AFM-like interaction when $V_{uu} > V_{ud}$ and FM-like interaction when $V_{uu} < V_{ud}$. As $V_{uu}/U > 0.2$ moves towards the repulsive region (from negative to positive), the amplitude of the $q=[\pi,\pi]$ susceptibility is greatly enhanced, while the $q=[0,0]$ mode is gradually suppressed. Conversely, in the $V_{uu}/U < 0.2$ case, the $q=[0,0]$ mode dominates while $q=[\pi,\pi]$ is suppressed. This behavior is consistent with is expected in FM and AFM-like interactions. Note that the $q=[\pi,0]$ and $q=[\pi,\pi]$ modes have comparable amplitudes only in the $V_{uu}< V_{ud}$ region, as shown in the inset of Fig.~\ref{fig:spin_kcut}(b).

Since we compute the full momentum resolution of $q$, we can obtain the spatial dependence of static $\chi_{s}(r)$. To do so, we evaluate $\chi_{s}(q)$ on an $L_{x} \times L_{y}$ external momentum grid and perform a discrete inverse Fourier transformation. Note  that the internal momentum resolution of $\chi_{s}(q)$ is continuously integrated out.  We take $L=36 \times 36$ such that the grid is sufficiently large enough to hold competing or coexisting commensurate and incommensurate magnetic ordering.  The three panels in Fig.~\ref{fig:FM_chi_r}(a-c) show false color plots of  $\chi_{s}$ in real space for $U/t=2,3$ and $4$. We see that centered around $r=0$, there is a short yet strong and robust ferromagnetic domain mediated by the $q=[0,0]$ mode. Furthermore, the domain size remains weakly dependent on $U$, spanning no more than $1 \sim 2$ lattice sites with correlation length $\xi_{0}< 1.2$. This indicates that a NN attractive  $V_{uu}$ interaction forces the system to create a localized parallel spin, resulting in a NN sized FM domain. Consequently, if next NN FM-like interaction or beyond is incorporated, the FM domain in real space can be expanded further.

When looking along the lattice axes, the FM domain decays to a quasi-long ranged stripe order, also known as a co-linear AFM order. This extended range suggests that the system transitions from localized magnetization to an itinerant magnetic order. The stripe exhibits significant strength along the $\hat{x}$ and $\hat{y}$ directions, as apparent in the $U=2$ case. Fig.~\ref{fig:FM_chi_r}(d) shows the y-axis cut of the three color plots. Here, we see that the FM domain has a large amplitude before decaying quickly to oscillatory behavior that persists over a longer range. The oscillatory amplitude, as shown by the inset figure, is faint  and is mediated by the wave vectors $q=[q_{x},0]$ and $q=[0,q_{y}]$ respectively.

Separated by FM and co-linear AFM quasi-domains is a checkerboard-like AFM domain located along the diagonal axes of the lattice. To assess phase separation between stripes and this checkerboard-like AFM domain, we take vertical y-axis cut at $r_{x}=6$ that is well beyond the FM domain in Fig.~\ref{fig:FM_chi_r}(e). The positive region corresponding to the first two lattices features the width of the stripes, beyond which we see an oscillatory region corresponding to the checkerboard AFM domain. Interestingly, increasing $U$ does not affect the qualitative feature and instead enhances the weight of the different domains. Finally, we see that soft magnetic fluctuations in the $q=[\pi,0]$ and $q=[\pi,\pi]$ modes seem sufficient to induce quasi-long-range order within our model, albeit about an order of magnitude smaller than shown in the $q=[0,0]$ mode. We have checked and verified that these features do not arise from finite size effects due to Fourier transformation on the discretized external momenta. 

\subsection{Fluctuating Magnetic orders} \label{sec:model}

 With the goal of understanding the spin susceptibility results in the previous section, we take a step back and propose a simple qualitative model that contains the salient features of the numerical susceptibility calculations.  This will allow us to see what implications a fully ordered system would have on different observables.  

\begin{figure}
    \centering
    \includegraphics[width=1\linewidth]{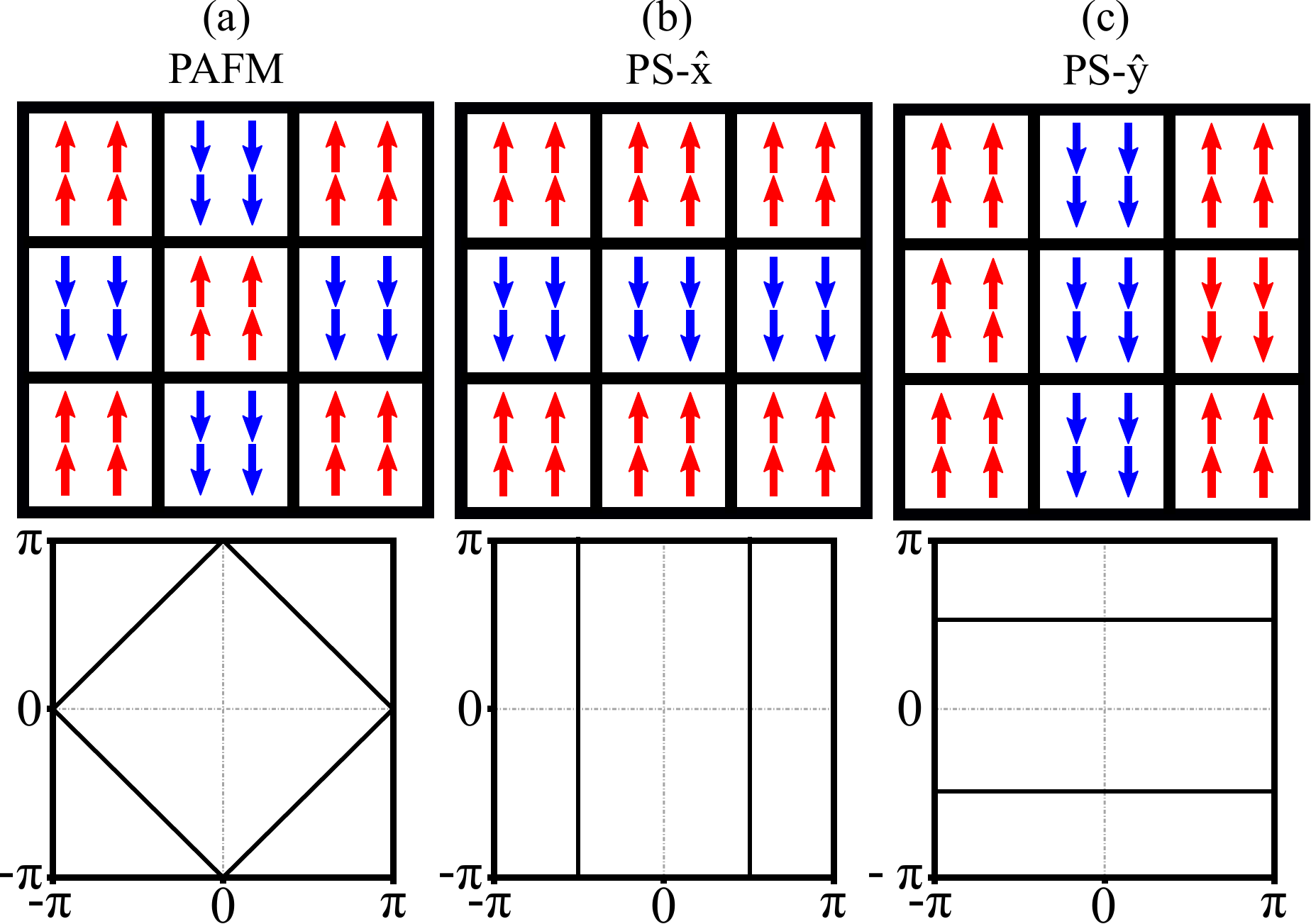}
    \caption{Schematics showing (a) plaquette antiferromagnetic (PAFM) ordering, (b) plaquette stripes (PS) in the $\hat{x}$-direction, and (c)  plaquette stripes (PS) in the $\hat{y}$-direction and their corresponding fermi surfaces (FS) by plotting Eq.~\ref{eq:PAFM} and Eq.~\ref{eq:PS}.}
    \label{schematics}
\end{figure}

Based on the spin susceptibility results in Fig.~\ref{fig:FM_chi_r}, we expect a short-ranged FM domain spanning a single site that eventually decays to form a quasi long-ranged co-linear or staggered AFM order.  We can classify such order into two types, plaquette AFM (PAFM) and doubly degenerate plaquette stripes (PS)\cite{Ruan2021,Gotze2013}. These plaquette structures consist of $2 \times 2$ sized sublattices with aligned spins within each plaquette, resulting in a net local magnetization while global magnetization remains zero. The FM plaquettes can either alternate in a staggered fashion forming PAFM or align in a stripe-like pattern along the two axes, forming a doubly degenerate PS configuration. Since the long ranged order formation is not possible in the thermodynamic limit at finite temperature, PAFM and PS should be considered as magnetic fluctuations rather than order formation. Fig.~\ref{schematics}(a-c) depicts the spin configuration of PAFM, and doubly degenerate PS along $\hat{x}$ and $\hat{y}$ in the top row, along with their corresponding Fermi surfaces in the bottom row. These two forms of magnetic fluctuation represent a hybrid of magnetic orders that are itinerant macroscopically while localized within the plaquette.  We further note that our calculations of PS order closely resemble the partially polarized itinerant ferromagnetic order of Nagaoka types obtained from density-matrix renormalization group (DMRG) calculations in the doped Hubbard model, where the $t$-hopping of a single electron on a doubly occupied site was shown to be enhanced \cite{Samajdar1,Samajdar2}.

Within each plaquette, the hopping is restricted between the aligned spins due to Pauli's exclusion principle which introduces strong kinetic frustration within the system. Due to the staggered arrangement of the plaquette in the PAFM case, hopping can occur between the NN plaquette via finite $t$, and no modification to band structure other than a reduction in bandwidth is expected. The effective dispersion of PAFM reads as 
\begin{equation}
    \bar{\epsilon} \approx -2[\cos(k_{x}) + \cos(k_y)]/\mathcal{D},
    \label{eq:PAFM}
\end{equation}
 where $\mathcal{D}$ is the  size of the $\mathcal{D} \times \mathcal{D}$ ferromagnetic plaquette and $(k_{x},k_{y})$ are the single particle momenta. The size $\mathcal{D}$ is determined from the  correlation length $\xi_{0}$ of $q=[0,0]$ mode where it scales as $\mathcal{D}=2\xi_{0}$. As the plaquette size grows, the system becomes increasingly localized, which further flattens the band structure. However, the Fermi surface of PAFM is expected to retain the nested feature of the original paramagnetic state for $U/t \to 0$ as shown on the bottom panel in Fig.~\ref{schematics}(a) and is independent of $\mathcal{D}$. On the other hand, PS-$\hat{x}$ and PS-$\hat{y}$ present a more interesting scenario where kinetic hopping is not only limited within the plaquette but also along the axis perpendicular to the stripe pattern. This magnetic anisotropy renders the quasi-particle dispersion of PS-$\hat{x}$ and PS-$\hat{y}$ as dispersionless along the x-axis and y-axis, respectively, yielding

\begin{equation}
    \bar{\epsilon}\approx
    \begin{cases}
     [2\cos(k_{y})]/\mathcal{D} & \text{for PS-$\hat{x}$},\\
     [2\cos(k_{x})]/\mathcal{D} & \text{for PS-$\hat{y}$}.
    \end{cases}
    \label{eq:PS}
\end{equation}
 Consequently, when sufficiently large PS fluctuations occur in the paramagnetic state, a single band is expected to split into two quasi-one dimensional bands, triggering a Lifshitz transition as indicated by the plotted Fermi surfaces in the bottom row of Fig.~\ref{schematics}(b,c). 

\subsection{Single-Particle excitations}
If the fluctuating PS and PAFM orders suggested by $\chi_{s}$ can be observed in single-particle excitations then we expect they will mimic the qualitative features of single particle properties identified in Sec.\ref{sec:model}, namely the compression of the bandwidth in Eq.(\ref{eq:PS}) and the shape of the Fermi surfaces in Fig.\ref{schematics}.  Starting from the non-interacting Green's function ($G_{0} = [\omega - \epsilon_{k}]^{-1}$) we compute diagrammatically the irreducible self energy ($\Sigma$) to fourth order from which the fully interacting Green's function ($G$) is obtained using the Dyson series $G^{-1} = G_{0}[1-\Sigma G_{0}]^{-1}$. The spectral function $A$ can be obtained from the imaginary part of $G$ on the real-axis via $A(k,\omega) = -\frac{1}{\pi}\text{Im}G(k,\omega)$ which leads to the usual Lorentzian form
\begin{equation}
A(k,\omega) =\frac{1}{\pi} \frac{-\text{Im} \Sigma(k,\omega)}{[\omega- (\epsilon_{k}+\text{Re}\Sigma(k,\omega))]^2 +[\text{Im} \Sigma(k,\omega)]^2}.
\end{equation}
Writing in this form emphasizes the role of $\text{Im} \Sigma(k,\omega)$ as the inverse lifetime and  $\tilde{\epsilon}(k) = \epsilon(k)+\text{Re}\Sigma(k,\omega)$ as the effective quasi particle dispersion.
\begin{figure}
    \centering
    \includegraphics[width=1\linewidth]{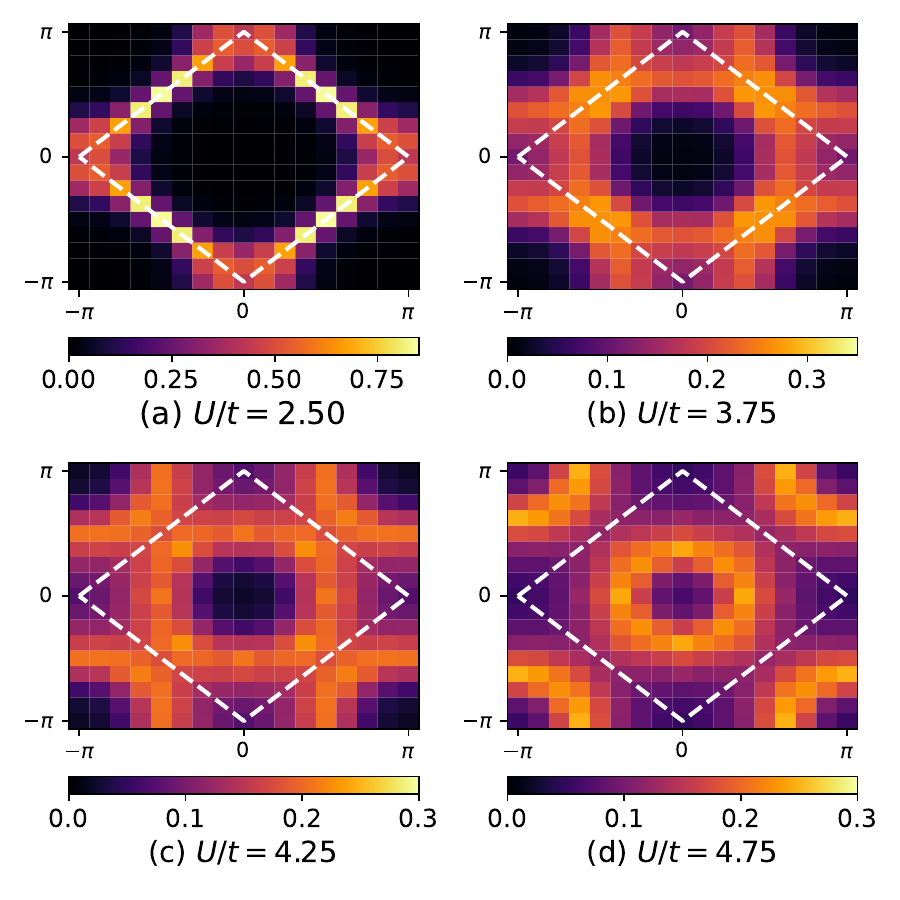}
    \caption{ (a-d)  Fermi surface at half filling plotted for four different values of onsite $U/t$ interaction and at $\beta=5$. The extended interaction parameters are fixed at the ratio of $V_{uu} = -0.2U$ and $V_{ud} = 0.2U$. The onset of the Lifshitz transition happens at $U=3.75t$. }
    \label{FM_fermi}
\end{figure}
Fig.~\ref{FM_fermi} presents the Fermi surface (FS) of the system at different interaction strengths. These were computed by calculating $A(k,\omega=0)$ perturbatively with fixed ratio of $V_{uu}/U=-0.2$ and $V_{ud}/U =0.2$ on an $L= 16 \times 16$ external momentum grid while the remaining internal momenta summations are continuous. Starting with $U=2.5t$, the FS retains the nested feature of the non-interacting FS marked by the dashed white line. However, some redistribution of states can be seen near the $k_{an}=(\pi,0)$ anti-nodal region, while the coherent spectral weight is confined around the $k_{n}=(\pi/2,\pi/2)$ nodal region, resembling pseudogap phases typical of the $V=0$ Hubbard model. The key distinction here is that the pseudogap feature arises from kinetic frustration within the short-ranged yet dominant FM fluctuations, while the AFM fluctuations remain suppressed.

Upon increasing the interaction strength to $U/t=3.75$, the states near $k_{an}$ are substantially suppressed, giving rise to fermi arcs. The renormalized FS presents significant modification compared to non-interacting FS topology, marking the onset of the Lifshiftz transition. At $U/t=4.25$, the system has already undergone a Lifshiftz transition and has formed two quasi-1D bands connected at $k_n$. The FS topology at this stage resembles the effective dispersion of the kinetically frustrated PS-$\hat{x}$ and PS-$\hat{y}$ magnetic orders in Eq.~\ref{eq:PS}. Lastly, for $U/t=4.75$, the spectral weight near $k_{n}$ is fully suppressed, rendering the FS fully disconnected. In this case, the FS topology is now that of a single, nearly circular hole pocket centered at $k=[0,0]$ with four electron holes at each corner of the Brillouin zone. Further, there is an overall increase in spectral weight in the hole and electron pockets compared to the $U/t=4.25$ case. We emphasize here that all the single particle calculations are at half-filling with $t^\prime=0$ and that the Lifshiftz transition is explicitly driven by the NN FM-like interaction for $V_{uu}< V_{ud}$. 
\begin{figure}
    \centering
    \includegraphics[width=1\linewidth]{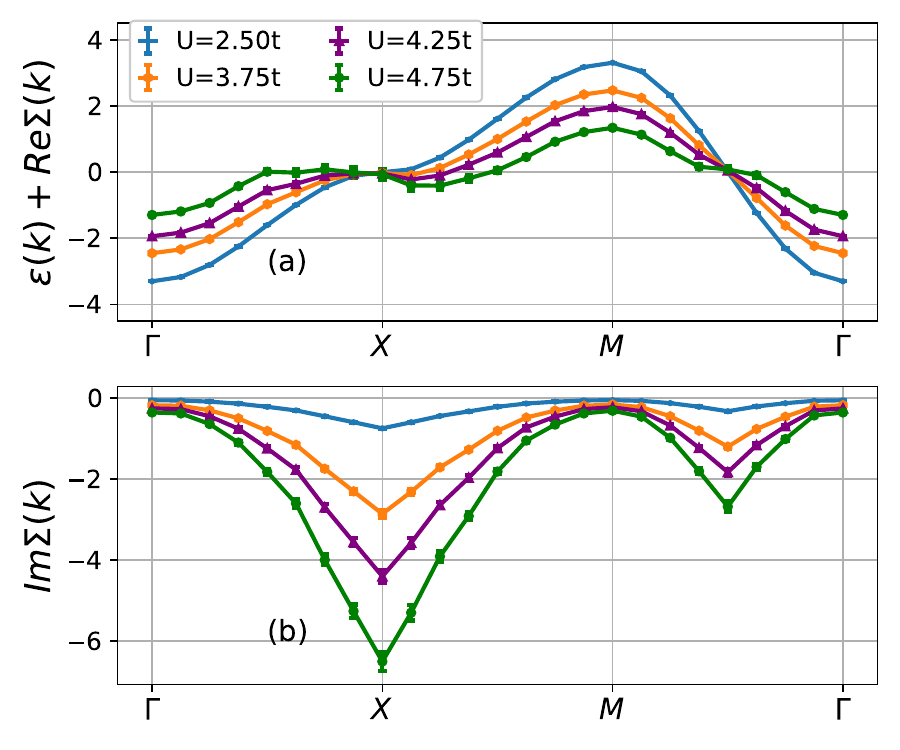}
    \caption{(a) Renormalized dispersion at $\omega=0$ calculated via $\Tilde{\epsilon_k}(\omega=0) = \epsilon_k + \text{Re} \Sigma(k,\omega=0)$. Lifshiftz transition occurs due to formation poles in low-energy excitation near the Fermi surface. (b) Corresponding $\text{Im}\Sigma(k,\omega=0)$ along the high symmetry cut in $k$ where   $\Gamma = [0,0]$, $X=[\pi,0]$ and $M=[\pi,\pi]$.}
    \label{FM_kcut}
\end{figure}

In order to understand the microscopic mechanism behind the Lifshiftz transition, we present in Fig.~\ref{FM_kcut} the low-lying single particle excitation, $\tilde{\epsilon}_{k}(\omega=0)=\epsilon_{k}+ \text{Re}\Sigma(k,\omega=0)$ and $\text{Im} \Sigma(k,\omega=0)$ along the high symmetry momentum cut. When $\tilde{\epsilon}_{k}(\omega=0)$ approaches zero, it corresponds to the pole in the spectral function, resulting in a peak in the FS. The width of the peak is controlled by $\text{Im}\Sigma(k,\omega)$. For sufficiently large $\text{Im}\Sigma(k,\omega)$, the spectral weight at the pole becomes incoherent, washing the states away. A Lifshiftz transition can occur in two mutually non-exclusive routes. First, a large $\text{Im}\Sigma(k,\omega=0)$ can disperse or wash away states, causing a change in the topology of FS. Second, highly anisotropic momentum-dependent $\text{Re}\Sigma(k,\omega=0)$ can produce additional poles and create a new coherent state on the FS. In Fig.~\ref{FM_kcut}(a), as $U/t$ increases, $\tilde{\epsilon}_{k}(\omega = 0)$ flattens across the entire momentum space. This flattening is particularly pronounced near the van Hove singularity (VHS) at $k_{an}$, where anisotropy is strongest. The additional poles formed along the $X$-$M$ line, combined with the loss of spectral weight at the $k_{an}$, results in the pseudogap phenomenon which is characterized by the Fermi arcs seen in Fig.~\ref{FM_fermi}(a,b). Similar anisotropy in the effective mass near the VHS is well documented in cuprates \cite{Katanin2004}. As $U/t$ increases further, the flattening of $\tilde{\epsilon}_{k}$ shifts the pole formation away from the VHS along the $X$-$\Gamma$ and $X$-$M$ lines, creating electron-like and hole-like states. Meanwhile, the original poles at $k_n$ and $k_{an}$ are progressively washed out due to the large increase in $\text{Im}\Sigma(k, \omega = 0)$, which destroys the original FS. At the same time, additional poles formed away from $k_{an}$ remain coherent, triggering a Lifshitz transition. This can be understood by noting that for a fixed value of $U/t$, as we move away from $k_{an}$, $\text{Im}\Sigma(\omega=0)$ decreases monotonically until it reaches $k_{n}$.  This also explains why hole and electron pockets see a sudden increase in spectral weight despite the insulating tendencies expected at $U/t=4.75$ in Fig.~\ref{FM_fermi}(d).  Consequently, we see that the interplay of $\tilde{\epsilon}_{k}(\omega = 0)$ and $\text{Im}\Sigma$, driven by $V_{uu}$, result in a Lifshitz transition, as observed in Fig.~\ref{FM_fermi}(c,d). 
\begin{figure}
    \centering
    \includegraphics[width=1\linewidth]{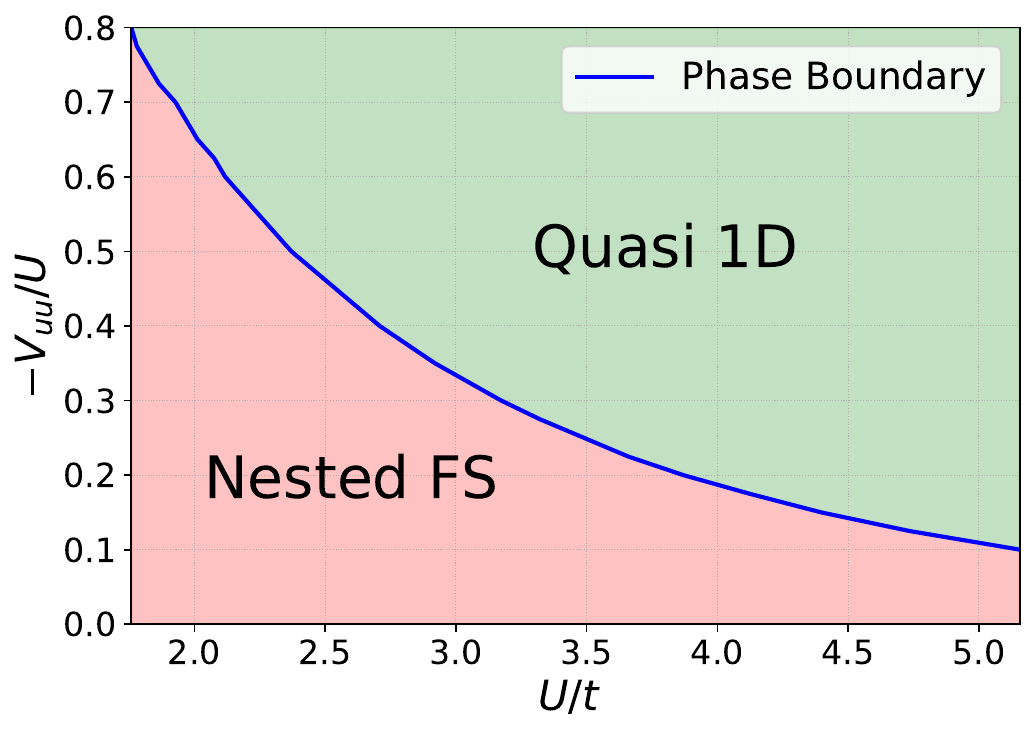}
    \caption{Phase boundary determined by the formation of at least two additional poles in $\Tilde{\epsilon}(k) = \epsilon(k) + \text{Re} \Sigma(k,\omega=0)$ along the $X \to M $ line. Here we set $V_{ud} = 0.2U$. Region marked by circled $(1)$ and $(2)$ represented nested FS and quasi-1D FS, respectively.} 
    \label{Phase}
\end{figure}

\begin{figure*}
 \centering
    \includegraphics[width=0.95\linewidth]{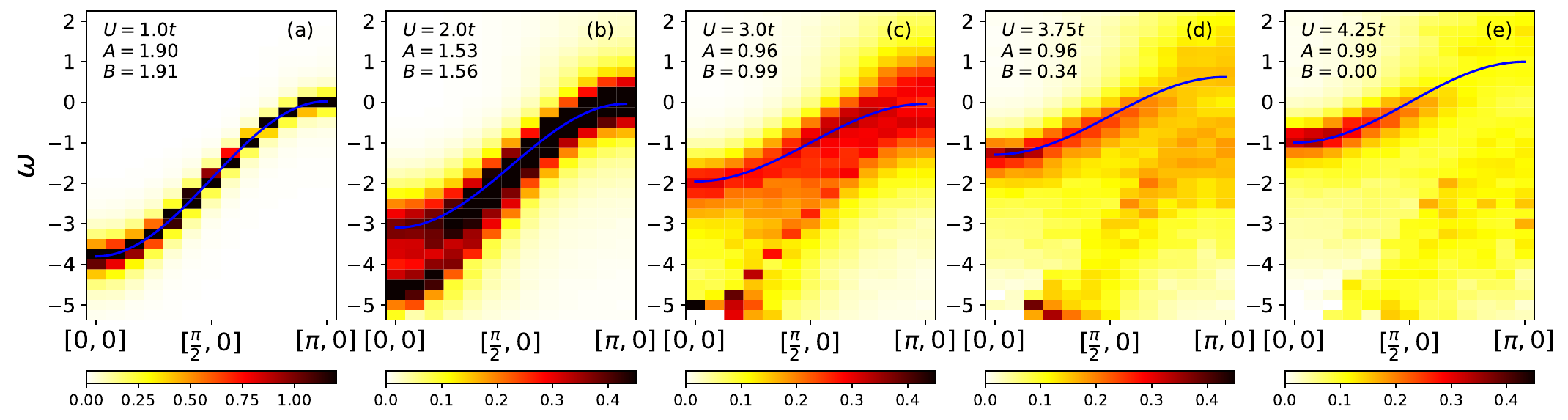}
    \caption{(a-e) Evolution  of spectral function near bottom of the band from $k=[0,0]$ to $k=[\pi,0]$ as a function of $U$. We utilized a fit equation of $\Tilde{\epsilon}=-[A\cos(k_x) + B\cos(k_y)]$ to upper brach that is splitting. Here we set $V_{ud} = 0.2U$ and $V_{uu} = -0.2U$ and use analytic continuation $i\omega \rightarrow \omega+i\gamma $ with $\gamma=0.2$. At $U=4.25$, we see effective band is that of quasi 1D with same $\bar{\epsilon}$ of PS-$\hat{y}$ as in Eq.~\ref{eq:PS} with $\mathcal{D}=2$. }
    \label{FM_band}
\end{figure*}
We have so far considered the simplest case of a FM-like NN interaction with equal amplitude but opposite sign, $V_{uu}/U=-0.2$ and $V_{ud}/U=0.2$, and have demonstrated both the quasi-1D effect and a Lifshitz transition arising in the intermediate coupling range of $U/t \geq 3.75$. However, it is possible to attain these effects even in the weak coupling limit by fine-tuning the $V_{uu}$ parameter, which is fully accessible through the computed multi-indexed coefficients of Eq.~\ref{eq:coeff}. Fig.~\ref{Phase} presents the phase boundary that marks the onset of the Lifshitz transition in the $V_{uu}/U - U$ phase space for a fixed $V_{ud}/U = 0.2$. The pink region corresponds to a squared FS with nested features, whereas the green region  represents a FS with two quasi-1D bands. The phase boundary is determined by the formation of at least two additional poles next to $k_{an}$ along the $X \to M$ line.  As $V_{uu}/U$ becomes increasingly attractive, the Lifshitz transition occurs at much lower $U/t$ values. For instance, Fig.~\ref{Phase} demonstrates when $V_{uu}/U = -0.2$, the Lifshitz transition is observed at $U/t = 3.75$. This is further corroborated by Fig.~\ref{FM_fermi}(b). When doubling the magnitude to $V_{uu}/U = -0.4$, the Lifshitz transition shifts to $U/t = 2.70$. This clearly indicates that flattening of $\tilde{\epsilon}_{k}(\omega = 0)$ and the corresponding pole formation is tied to how `ferromagnetic' the NN spin polarized interactions are.

We study the evolution of the band along the $k = [0, 0]$ to $k = [\pi, 0]$ momentum cut in Fig.~\ref{FM_band}(a-e) as a function of $U/t$.  The effective quasi-particle band is fit to the equation $\Tilde{\epsilon} = -[A\cos(k_x) + B\cos(k_y)]$
in order to extract the half the bandwidth $(A + B)$ and assign the dominant magnetic fluctuation as PAFM or PS. When fluctuations in PS-$\hat{y}$ set in, we expect the hopping along the $y$-axis to be dramatically reduced and characterized by a significant reduction in $B$ compared to $A$. At $U/t = 1$,  renormalization effects are weak, and the band resembles that of a non-interacting system. Increasing to $U/t = 2$, we observe a splitting of the band near the $\Gamma$ point, with one branch approaching the FS and the other away from it. We fit the former and find that the weight of $A + B$ is reduced, with $A \approx B$ indicating no kinetic frustration along a particular axis.  At $U/t = 3$, we find the bandwidth has reduced by half, but the weight of $A$ remains comparable to $B$. Setting $U/t = 3.75$, near the onset of the Lifshitz transition, we observe a complete splitting of the band. The lower branch exhibits a heavy loss of spectral weight, while the upper branch remains coherent. Fitting the upper branch reveals a further reduction in bandwidth, driven primarily by a decrease in $B$, while $A$ remains unchanged from the $U/t = 3$ case. Finally, at $U/t = 4.25$, where the FS consists of two decoupled 1D bands, the weight of $B$ approaches zero and the system has transitioned to a quasi-1D state driven by dominant fluctuation in PS-$\hat{y}$ order. From the fits, we see that $A \approx 1.0$ while $B=0$, which exactly match $\bar{\epsilon}$ of PS-$\hat{y}$ fluctuations in Eq.~\ref{eq:PS} with $\mathcal{D}=2$. Note that a similar quasi-1D effect can be observed when looking along $k = [0, 0]$ to $k = [0, \pi]$ , but instead of $B\rightarrow0$, the largest suppression is seen in $A$ when PS-$\hat{x}$ is the dominant fluctuations.

\begin{figure}
    \centering
    \includegraphics[width=1\linewidth]{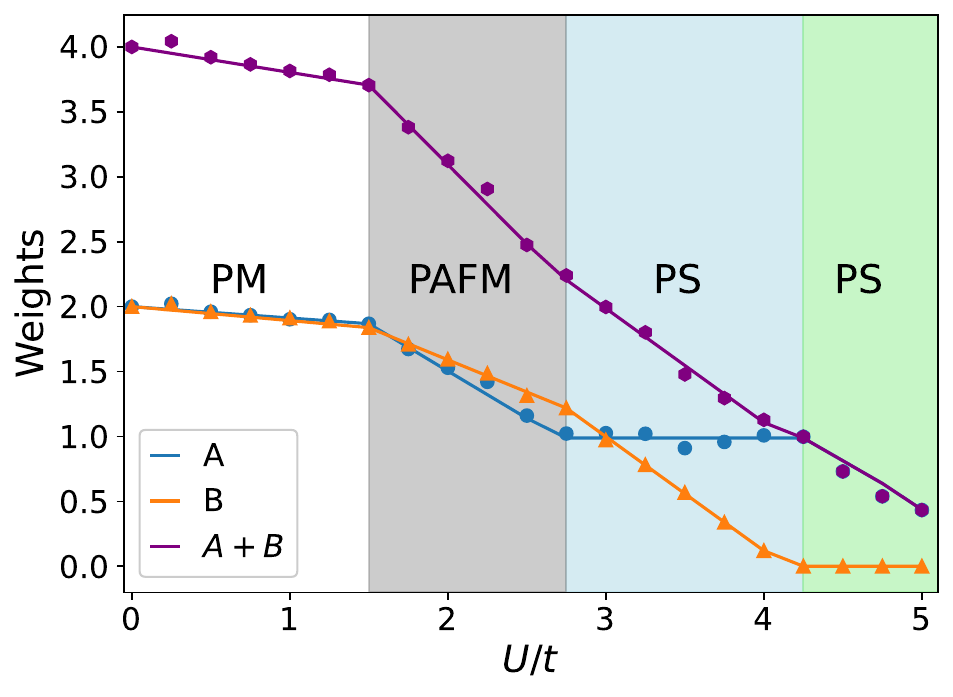}
    \caption{Weight of $A$,$B$ and the half the bandwidth $A+B$ extracted from the fit equation   $\Tilde{\epsilon}=-[A\cos(k_x) + B\cos(k_y)]$ utilized in Fig.~\ref{FM_band} plotted as a continuous function of $U/t$. Based on the fits, we distinguish the region into three possible forms of dominant magnetic fluctuation: Paramagnetic (PM), plaquette AFM (PAFM), and plaquette stripe (PS). }
    \label{fits}
\end{figure}
Fig.~\ref{fits} shows the evolution of the fitted parameters $A$ and $B$ and the corresponding half-bandwidth $(A+B)$ as a continuous function of $U/t$. In the range of $0<U/t<1.5$, marked by white shaded region, we see a steady and slow decline in the magnitude of both $A$ and $B$. Here, the kinetic frustration is minimal and the system can be described as in a paramagnetic (PM) state. From the region $1.5<U/t< 5$, we see a steep but linear decline in the bandwidth. The weight of $A$ and $B$ contribute differently to the reduction of bandwidth. In the range of $1.5<U/t<2.75$ (grey region), the weight of $A$ declines slightly faster than $B$, but is largely equal, indicating no hindrance of kinetic hopping along a particular axis. This region is characterized by dominant fluctuations in PAFM order. Within the region  $2.75<U/t<4.25$ (blue region), we see that fluctuation in $A$ is minimum while $B$ declines rapidly to zero, indicating a strong kinetic frustration along the $y$-axis. Thus, this region is characterized by dominant fluctuation in the PS and indicates the onset of the Lifshitz transition is due to gradual transformation to quasi-1D bands. Finally, in $4.25<U/t<5.0$, only the weight of $A$ falls while $B$ remains static and equal to zero, giving rise to hole and electron pockets seen in the FS.

\subsection{Leading Pairing symmetry }
In order to identify the leading pairing symmetry within the system consisting of localized $2 \times 2$ FM plaquettes with $V_{uu} < V_{ud}$, we have also calculated the uniform pairing susceptibility in the normal state via
\begin{equation}
    P^{g}_{\sigma,\sigma^\prime} = \int_{0}^{\beta} d\tau \langle \Delta_{\sigma\sigma^{\prime}}^{\dagger g}(\tau) \Delta_{\sigma\sigma^{\prime}}^{g}(0) \rangle,
    \label{}
\end{equation}
where$\Delta_{\sigma\sigma^{\prime}}^{\dagger g} = \sum_{\mathbf{k}}g(k)c^{\dagger}_{\mathbf{k}\sigma}c^{\dagger}_{\mathbf{-k}\sigma^{\prime}}$ is the superconducting order parameter projected onto the symmetry factor $g$. We diagrammatically expand $P^{g}_{\sigma,\sigma^\prime}$ up to fourth order and include only the vertex diagrams to compute the correlated pairing response. We consider NN and even frequency pairing of three types: $p_{x}+ip_{y} \to \sin(k_{x})+i \sin(k_{y})$, $d_{x^2-y^2} \to \cos(k_{x}) - \cos(k_{y})$, and $s_{x^2+y^2} \to \cos(k_{x}) + \cos(k_{y})$ \cite{Sigrist}. 

The time reversal symmetry breaking $p_{x}+ip_{y}$ pairing corresponds to a triplet symmetry with even spatial parity, allowing Cooper pairs with three possible spin projections. The $S_z = 0$ state has a symmetric spin mixed configuration, $\frac{1}{\sqrt{2}} \big(|\uparrow \downarrow \rangle + |\downarrow \uparrow \rangle\big)$. The $S_z = +1$ and $S_z = -1$ states involve equal-spin pairing, represented as $|\uparrow \uparrow \rangle$ and $|\downarrow \downarrow \rangle$, respectively, and are degenerate. In our system, the triplet equal-spin $p_{x}+ip_{y}$ pairing ($S_z = \pm1$) emerges as a natural candidate for the leading pairing symmetry due to short ranged FM plaquettes and the attractive interaction between NN equal spins (i.e., $V_{uu} < 0$). From perturbation theory, the singlet $d_{x^2-y^2}$ pairing is the leading symmetry at half-filling when $V=0$ \cite{me2023}. Meanwhile, the singlet $s_{x^2+y^2}$ pairing becomes relevant in the region where the FS Lifshiftz transition to electron and hole pockets. This is due to the gap function's sign-changing property between the electron and hole pockets located at the corners and center of the Brillouin zone, which minimizes the repulsive interaction ($V_{ud}/U=0.2$) in a singlet Cooper pair.

\begin{figure}[h]
    \centering
    \includegraphics[width=1\linewidth]{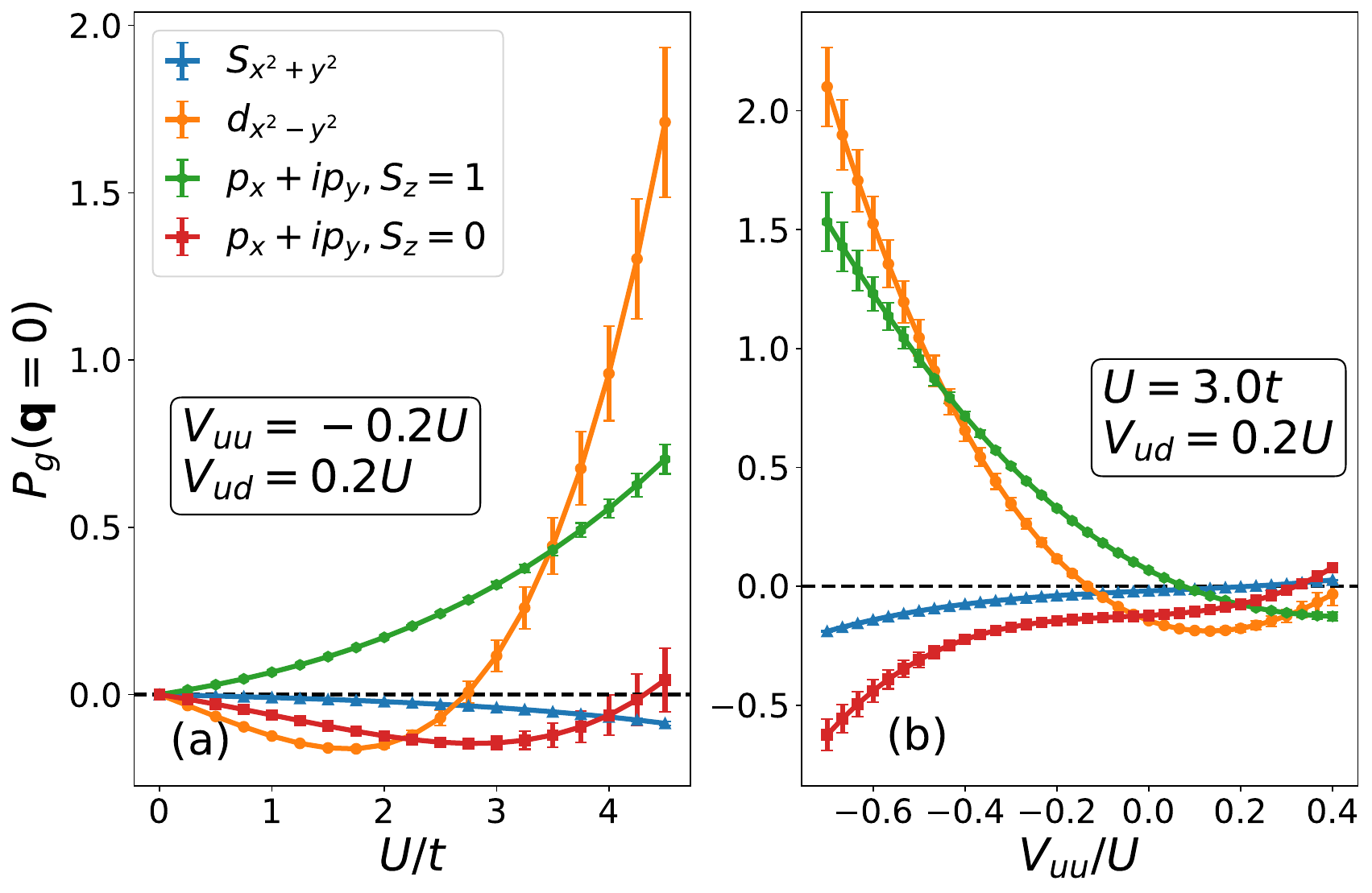}
    \caption{(a) Uniform pairing susceptibility ($P_{g}$) as a function of $U$ for different symmetry channel with fixed $V_{uu}/U=-0.2$ and $V_{ud}/U=0.2$ (b) $P_{g}$ as a function of $V_{uu}/u$ fixed $U/t=3.0$ and $V_{ud}/U=0.2$. }
    \label{pairing}
\end{figure}

Fig.~\ref{pairing}(a) presents the evolution of $P^{g}_{\sigma\sigma^{\prime}}(q=0)$ as a function of $U$, with fixed $V_{uu}/U = -0.2$ and $V_{ud}/U = 0.2$ at $\beta t = 5$. We observe that the triplet $S_z = 1$ $p$-wave pairing is attractive in the $U \to 0^{+}$ limit, while other symmetries remain repulsive. The equal spin $p$-wave pairing increases monotonically, in contrast to unequal-spin $p$-wave pairing which remains repulsive throughout with uncertainty. Around $U/t = 2.75$, the $d_{x^2-y^2}$ pairing changes sign and eventually becomes the leading pairing symmetry at $U/t = 3.5$.  Notably, the onset of the quasi-1D band formation identified in Fig.~\ref{fits} coincides with the region where $d_{x^2-y^2}$ changes sign. However, the $s_{x^2+y^2}$ pairing remains repulsive throughout the entire range of $U/t$ probed, despite the Lifshitz transition. Finally, by fixing $U/t = 3$ and $V_{ud} = 0.6$, we study the pairing response as a function of $V_{uu}$ in Fig.~\ref{pairing}(b).  Recall that increasing the amplitude
of $V_{uu}$ in the attractive regime enhances fluctuations in the $q = [0, 0]$
and $q = [\pi, 0]$ spin modes thereby enhancing PS magnetic fluctuations. Here, we see that $S_{z}= \pm1$ $p$-wave and singlet $d_{x^2-y^2}$ pairing also receives similar enhancement as the  
 attractive strength of $V_{uu}$ increases. 
 
 Now a clear picture emerges of intra-plaquette $p$-wave Cooper pair formation between adjacent sites with equal spin in the weak coupling limit, mediated by an attractive NN interaction and strong FM $q=[0,0]$ fluctuation. In the intermediate coupling regime, the triplet $p$-wave pairing competes with the singlet $d_{x^2-y^2}$ channel, which involves inter-plaquette bonding between unequal spins at adjacent sites, mediated by a repulsive NN interaction and soft fluctuations in co-linear $q=[\pi,0]$ AFM modes.

\section{Discussion and Outlook}
Motivated by the recent findings that a renormalized local $U/t$ interaction gives rise to spin-dependent non-local interactions \cite{gazon2023}, we have explicitly incorporated spin-differentiated NN interactions into the standard Hubbard Hamiltonian. We have shown that when NN interactions are FM-like (i.e., $V_{uu}<0$, $V_{ud}>0$), the Hubbard model displays a rich interplay of various magnetic fluctuations, namely plaquette AFM and doubly degenerate plaquette stripes with short-ranged $2 \times 2$ FM domains. Although such magnetic ordering possesses no global net magnetization, localized FM domains are rare and have been of particular interest with relevance to Nagaoka-type ferromagnetism. The induced FM domains introduce kinetic frustration in the system, leading to a monotonic but linear decrease in bandwidth with $U/t$. We see that FM domain formations are not of Nagaoka-type despite the superficial similarities, but rather arise as a mechanism to minimize the potential energy associated with spin polarized interactions that favor the alignment of equal spins. Kinetic frustration only acts as a secondary effect to stabilize the FM domains. We further observe that this FM-induced frustration creates a pseudogap, followed by a complete suppression of the nested Fermi surface and the emergence of two quasi-1D bands with hole and electron pockets. We emphasize that these can seen from bare perturbative expansion without
resorting to the mean-field approaches. While the flattening of bands coincides with the formation of a FM plaquette, the causal effect of the FM domain is reversed compared to flat band ferromagnetism. It was previously shown that the formation of a flat or nearly flat band near the FS under the presence of finite repulsive $U/t$ yields a ferromagnetic ground state \cite{Tasaki_1998}. However, within our model, the creation of localized FM domains by $V_{uu} < V_{ud}$ enhances $q=[0,0]$ spin excitation, which then mediates the gradual reduction in the bandwidth. 

We have also studied more naturally occurring AFM-like NN with $V_{uu}/U=0.2, V_{ud}/U =-0.2$ interactions, as shown to arise from renormalized $U/t$ in Ref.~\cite{gazon2023}. We have found that it leads to a massive enhancement of the $q=[\pi,\pi]$ spin excitations by an order of magnitude compared to $V=0$ or non-polarized $V_{uu}=V_{ud}$ case for a fixed $U$. We see further shifts of the metal-insulator transition to lower interaction strength of $U\approx 2$, followed by a large enhancement in $d_{x^2-y^2}$-wave pairing response. However, no signs of the Lifshitz transition or bandwidth reduction in $\bar{\epsilon}$ is seen. We see that introducing spin-polarized AFM-like NN interactions at a small $U/t$ limit replicates the physics of the strong coupling limit. Thus, our results underline the key role spin-differentiated interactions may play, which is often ignored. Furthermore, the model presented paves the way for an interaction-driven approach to reconstruct bands and FS, create localized ferromagnetic domains, and other collective magnetic phases without the need for explicit modification to single-particle hopping. Such spin-differentiated non-local interactions could potentially be attainable in optical lattice experiments, given that relevant phases occur at high temperatures \cite{O_Hague2021,O_Su2023}. Formation of spin and charge density waves in Hubbard model remains an active area of continued interest for cold-atom experiments \cite{Bourgund2025,O_Mazurenko,O_Xu}.

We have also demonstrated the condition under which equal-spin triplet pairing in the bulk may emerge by computing the pairing response. First, this requires the creation of short-ranged FM domains. This can be done by inducing a competition between strong FM fluctuations with shorter wavelength competing with collinear or checkerboard AFM with longer wavelength, thereby creating strong fluctuation in PS and PAFM magnetic ordering. Second, effective attractive interaction between neighboring sites within creates a potential for pairing between neighboring parallel spins within the plaquette. In fact, strong anomalous attractive NN interaction has been reported in cuprate chains via a phononic mechanism \cite{Chen}. Subsequent theoretical studies have presented the possibility for enhanced triplet $p$-wave pairing \cite{Qu}. We speculate that increasing the plaquette size may enable FM plaquettes to host more exotic long-range pairings, such as equal spin triplet $f$-wave, which can likely be achieved by further FM-like next-NN coupling or beyond. Thus, our model can serve as a minimal single-band model under which equal spin triplet $p_{x}+ip_{y}$ wave pairing emerges more intrinsically.

We have presented our calculations at a nominal temperature limit of $\beta t = 5$, ensuring that the fourth-order expansion has minimal higher-order corrections. Within this temperature range, it is well established that key finite-range orders such as AFM fluctuations, incommensurate charge ordering, and pair-field fluctuations—compete or coexist in the $V = 0$ Hubbard model at the thermodynamic limit \cite{Schafer2021,me2023,McNiven2022,Dong2020,Maier2019}. Notably, the key results in our study occur within parameter regimes accessible to a wide range of numerical techniques, including both perturbative and non-perturbative methods such as quantum Monte Carlo, dynamical cluster approximation, and dual-fermion approaches. At lower temperatures and stronger coupling regimes, non-perturbative methods may be particularly well suited but may suffer from spurious continuous phase transitions associated with finite-size effects. However, this does not occur within our approach. In subsequent studies, using either a perturbative or non-perturbative approach, it would be interesting to quantify the role of doping and next-NN  hopping on magnetic ordering and the leading pairing function within our model. In particular, we expect next-NN hopping or beyond to alleviate kinetic frustration along some preferential lattice axes. This can be exploited to further engineer the Fermi surface and tune the location of the van Hove singularities.

\section{Acknowledgement}
We acknowledge the support of the Natural Sciences and Engineering Research Council of Canada (NSERC) RGPIN-2022-03882.  Computational resources were provided by ACENET and the Digital Research Alliance of Canada.

\bibliography{refs.bib}
\end{document}